# Towards a Graph-Based Approach for Web Services Composition

Chaker BEN MAHMOUD[1], Fathia BETTAHAR[2], Hajer ABDERRAHIM[3] and Houda SAIDI[4]

[1]ENIG, University of Gabès
Gabès, Tunisia

[2]ISIMG, University of Gabès
Gabès, Tunisia

[3]ISIMG, University of Gabès
Gabès, Tunisia

[4]ISIMG, University of Gabès
Gabès, Tunisia

**Abstract**
Nowadays, Web services (WS) remain a main actor in the implementation of distributed applications. They represent a new promising paradigm for the development, deployment and integration of Internet applications. The aim of Web services composition is to use the skills of several departments to resolve any problem that cannot be solved individually. The result of this composition is a compound of Web services that define how they will be used. In this paper, we propose an approach for automatic web services composition based on the concepts of directed graphs for the representation and description of Web services, and the ordering of web services compound execution. In this context, the user query, defined by a set of inputs and outputs, can be viewed as a directed graph composed of Web services.

***Keywords:*** *Web Services, Automatic composition, WSDL, service-oriented architecture, Theory of graphs.*

## 1. Introduction

Web services provide a new way to develop distributed and dynamic applications. They are considered to be a good solution for interoperability during data exchange between heterogeneous applications within an organization. One of the most important advantages of the Web service is reuse. In fact, Web services are conceptually limited to relatively simple features which are modeled by a collection of operations.

The use and composition of Web services to solve problems remains a difficult task to achieve, however. Web services composition refers to the process of creating a composite service with new functionality from existing relatively simple Web services. This process includes discovery, integration and execution of Web services in a specific order to meet an identified need.

Intense research activities have been conducted in this area in order to achieve correct web services composition. In fact, composition of Web services is not just simple grouping, but rather a composition in which the web services are ordered according to the relations between their semantics. These are usually provided by different organizations independently from any execution context. Since each organization has its own working rules, Web services should be treated as strictly autonomous units.

## 2. Related work

Several approaches have been developed for Web services composition. Only few used the concept of graph theory. In what follows, we present three approaches closely related to the one presented in this paper.

Elmaghraoui *et al.* presented in [1] a solution for optimizing the computational effort in Web services composition. This approach is based on graph theory. It consists in modeling the relationship between the involved semantic Web services in a directed graph, and calculating the shortest path by using an extended version of the Floyd-Warshall algorithm. This optimization approach is based on two pillars: i) the first is defining the semantic relationships between the available Web services using an directed graph called Service Composition Graph (SCG), and ii) the second is applying a graph search algorithm to calculate the shortest paths between all nodes. Finally, the results of the algorithm are stored in a matrix called the Shortest Predecessor Matrix (SPM).

Hashemian *et al.* [2] has created composite Web services using a graph search algorithm based on input/output dependencies between Web services. In fact, the composite Web services are presented as a dependency graph built using input-ouptut requirements of available elementary web services. A dependency graph $G = (V, E)$ contains information about the existing Web services in the repository as well as their input/output. The set of nodes V represents the actions or statements on inputs/outputs included in the list of inputs/ outputs. There is a directed edge from node $v_x$ to node $v_y$ in the graph (where $v_x, v_y \in$ V) if and only if there exists at least one dependency $v_x \rightarrow v_y$ in the list of dependencies between inputs and outputs. Each edge in E is a set that contains all web services in the repository having that dependency in one of their dependency sets. This algorithm resolved the composition problem in two steps: i) find Web services that can potentially participate in the composition, ii) find the composition based on these Web services. The author considered the dependencies between input and output parameters without considering the semantic functions, so they cannot guarantee that the generated composite service correctly provides the requested functionality.

Samuel *et al.* presented in [3] a composition technique based on weighted planning graph in which the composition can be found in polynomial dynamic time and in heterogeneous environment. The author conceived the composition problem as a problem of generating the required outputs from the given inputs. Therefore, the order of actions is not important, except that he supposed that the inputs arrive before outputs. Many information systems fall into this category where the inputs and outputs can be retrieved from different WSDL files. It uses a special graph structure called dependency graph to construct an index of available web services and their input/output information. This graph can be considered as a model for the repository specification, because it is accessible by the composition planner in response to a request. The planning graph is a layered directed graph. The vertices can be of two types. The first is the collection of propositions (pre-and post-conditions of Web services), called P and the second is the actions (set of Web services), called A. The edges connect one layer to another. The quality attributes can be assigned on the edges as weights. After the construction of the planning graph for composition, it applies the non-functional quality parameters to find a better composition scheme.

# 3. Model of Web services composition-based graph

The automatic composition of Web services is a complex task. Indeed, the use of Web services is limited, firstly because they perform a specific task. On the other hand, the structure and the availability of these services are not stable because of the exponential evolution of the Web. In this context, we propose a system for automatic composition of Web services. To achieve this goal, we consider that a solution for composition must be engaged from the Web services discovery to the interaction with users. From a user perspective, once the query is defined, the system commits to identify Web services necessary to satisfy the problem. In order to achieve these objectives, we propose a graph-based model for automatically composing Web services.

## 3.1 Principle

We assume that Web services are defined by a set of inputs and outputs describing their semantics. The user defines his needs in terms of inputs/outputs with a textual description. These needs must be validated with respect to the domain ontology (concept or term). The proposed system selects the Web service that is the most adapted to the given user inputs and uses the results to continue to meet the goal (all outputs). The resulting Web service can then be viewed as a Web services-based graph constructed according to input/output similarity. This approach covers the following features: service composition, discovery, execution and publication of the composite service.

## 3.2 Web service modeling

Each Web service contains different operations and is defined by its name, parameter, and state of the world in which it operates. In this work, we assume that each Web service consists in a single operation. For sake of simplicity, we use operation and Web service terms interchangeably.

We propose the following formalism:

*WebService (Parameters, State-of-the-world)*      (1)

This representation allows us to introduce a Web service as an entity that is fully defined by its parameters ("PARAMETERS ") and the state of the world ("State-of-the-world ") where it acts. Parameters are represented by the inputs and outputs of Web service, and the state of the world is represented by its preconditions and its effects.

*WebService (Inputs, Outputs, Pre-conditions, effects)*   (2)

Where, Inputs, Outputs $\subset$ Parameters and Pre-Conditions, Effects $\subset$ State-of-the-world.

This second representation was used to introduce a web service as an entity capable of producing one or more concrete results based on inputs/outputs requirements. Pre-conditions provide information about the state in which the world must come before the invocation of a service. The effects indicate the state of the world after the invocation of the service.

### 3.2.1 Conditions on web service's inputs and outputs

The inputs and outputs of a Web service should be able identifiable by a concept defined within a well-established ontology. Therefore, the different parameters of a Web service can be represented by instances of concepts belonging to different ontologies.

For example, the web service "Find_Doctor" uses a single input parameter (denoted City_Name) represented by the concept of "CITY", belonging to the ontology CNAMOnto.

So "Find_Doctor" requires an instance of the concept "CITY". The only output parameter (Doctor) is represented by an instance of the concept DOCTOR present in the ontology CNAMOnto. "Find_Doctor" will return an instance of the concept "Doctor" if the preconditions are validated. So the inputs and outputs of a web service are clearly defined in terms of concepts of a specialized ontology and for the attributes (e.g.CIN of any person or other).

3.2.2 Conditions on the state of the world (preconditions and effects) of the web service

In order to facilitate reasoning about preconditions and effects, we present them using the first-order predicate logic. Indeed, the preconditions and effects of web services are used to estimate the state of the world in a given situation. It is therefore essential to be able to reason with these world estimators.

So we adopt the following formalism to define the state of the world (preconditions and effects) of a web service:

*PreCondition(WS,PC1,...,PCn)←Valid(PC1)∧...∧Valid(PCn)*   (3)

*Effect (WebService,E1, ...,En) ← E1∧ ... ∧En*   (4)

In fact, a web service is defined by its parameters, but also the states of the world through which it passes. It is therefore necessary to integrate the pre-conditions and effects in the definition of a web service. For example, the web service "Find_Doctor" said two predicates P1 (pre-conditions on the service) and E1 (effects on the service).

*P1 (Find_Doctor, CITY) ← Exist (DOCTOR) ∧ Valid(CITY)*
*E1 (Find_Doctor, DOCTOR) ← List (DOCTOR)*

3.3 Operation of the automatic web services composition

3.3.1 Composition module

In this phase, we proceed to the automatic composition of web services. We focus on the operational aspects of web services and we concentrate particularly on the input parameters and output web services. An interaction graph of web services is represented as an oriented graph in which the vertices represent the set of web services and links materialize the flow of information between two web services.

To represent the interactions between a set of web services in the form of oriented graph, vertices can be defined using levels of details (parameter, service).

In an oriented graph, whose vertices are web services, links represent the common parameters (input / output) that allow web services interact.

Let A be a web service described by WSA ($I_A$, $O_A$, $Pre_A$, $Ef_A$), where $I_A$ is the set of inputs, $O_A$ the set of outputs, $Pre_A$ refers to preconditions and $Ef_A$ refers the effects.

In order to create a link between a source web service A described by ($I_A$, $O_A$, $Pre_A$, $Ef_A$) and a target Web service described by B ($I_B$, $O_B$, $Pre_B$, $Ef_B$), the number of output parameters $O_A$ of web service A must be greater than or equal to the number of input parameters $I_B$ of web service

B. In this context, two cases may arise: complete relation or partial relation:
- *Complete relation*: if and only if, for each input parameter of the target web service B, there is an output parameter similar in the web service A.
- *Partial relation*: there are at least one output parameter of the source web service similar to an input parameter of the target web service

**Process of building the graph composition**

In our approach to composition, the user request passes through several processing stages before constructing the graph composition. The resolution of this problem of composition identifies the resolution of a goal (outputs) described by the user's query.

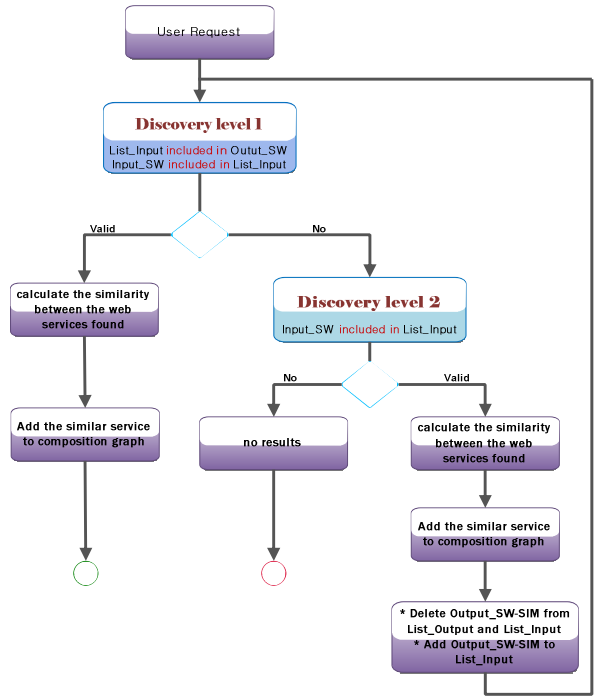

Fig. 1  Process of building graph composition

Example:
Let a set of web services declared as follows:
   WS1 ({a,b},{c,d,f},{P1},{EF1,EF2}) ,
   WS2 ({c},{m,k},{P2},{∅}) ,
   WS3 ({w,m},{t},{P3,P4},{EF3}) ,
   WS4 ({k,d,i},{p},{P5},{EF4}) ,
   WS5 ({f},{i,g},{P6},{EF5}) ,
   WS6 ({h,g,n},{y,q},{P7},{EF5}) ,
   WS7 ({a},{f},{P8},{EF}) ,
   WS8 ({t},{z,g},{P9},{∅})
And the request of the user is defined as follows:
ReqUti ({a,b,w},{t,p})

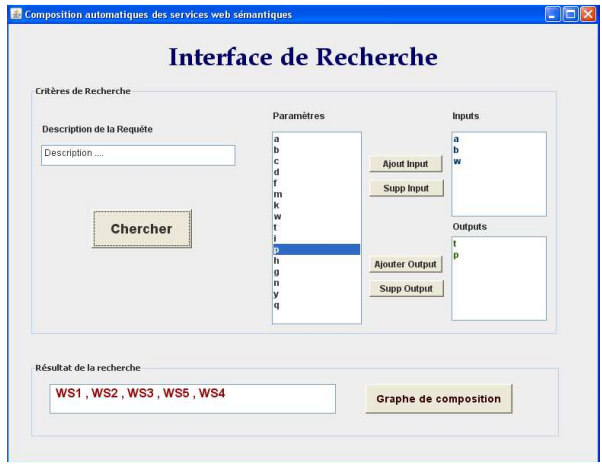

Fig. 2  Search Interface

The graph composition result in the execution of the user query is as follows:

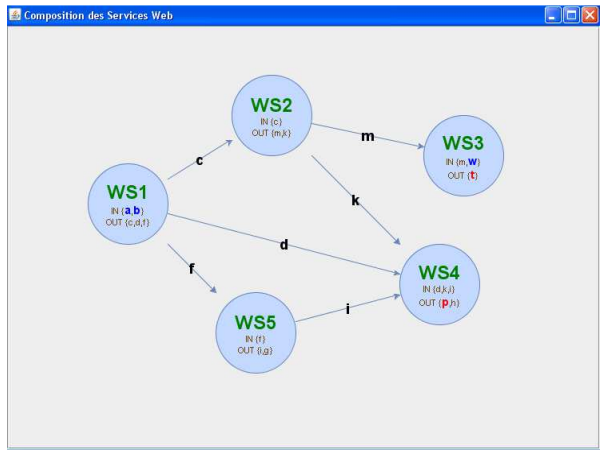

Fig. 3  Result of the composition (composition graph)

3.3.2 Discovery module

This module searches for a list of web services that meets the requirements expressed by the composition module.

The discovery module uses the properties of web services in order to find the ones that best respond to a query. In the discovery process, our module seeks the similarities between the query definition parameters and the web services ones published in registries.

The similarity calculation has a great influence on the search of web services for discovery and composition. For the discovery, the similarity calculation is based on textual descriptions, on the input and output parameters, and the state of the world of web services. For the composition, the similarity calculation is applied to the output parameters of the first service compared and on the input parameters of the second service compared.

Moreover, according to the nature of objects to compare, the similarity can be broken down into syntactic similarity or semantic similarity. Note that the matching semantics can be used on syntactic descriptions by enriching the descriptions for the treatment. Various solutions have been proposed in the literature like the use of tools such as lexical database WordNet [4] or methods such as latent semantic analysis [5].

**Syntactic similarity**

Syntactic similarity compares parameters from their respective orthographies. Two distinct similarities can be distinguished, the approximate similarity and the equal one.
- The equal similarity uses the strict syntactic equivalence. Two objects are said to be similar if and only if they have the same orthography.
- The approximate similarity uses distance functions $d(x,y)$ to quantify the similarity between two character strings x and y. If the distance between two objects is above a certain threshold, these objects are said similar.

**Semantic similarity**

For comparing the outputs of a request to the outputs of a published service, four degrees of matching are used [6]:
- *Exact matching*: select a web service if he corresponds exactly at the request (request = Service) that is to say, the inputs and outputs of the request are equal to the inputs and outputs of the web service.
- *Plug-in matching*: returns a web service if he includes a request (request < Service) that is to say, the input of the request includes the inputs of the web service and outputs of the request are subsumed by the outputs of the web service. In this case, the web service is a set that generalizes the request.
- *Subsumes matching*: returns a Web service, if he included in a query (request > Service). In this case, the service does not completely satisfy the request. This service may be used to achieve partially the purpose of the request. One or more additional services may need to be used to meet all the goals of the user.
- *Fail matching*: returns false if no match between the query and service.

In terms of satisfaction of the request, the semantic matching degrees can be ordered according to a scale of preference as follows:

*Exact > Plugin > Subsumes > Fail*

In our approach, the discovery of web services is to find links and semantic correspondences between the parameters of the request with Web Services. This discovery is essentially based on the parameters and the state of the world of web services.

In this context, the similarity can be divided into two parts: parameters similarity (Input and Output) and similarity of state of the world (Pre-condition and Effect).

The system measures the similarity of the parameters (input and output) by attributing a score for each mode of matching: Exact (score=3), Plug-In (score=2), Subsumes (score=1), Fail (score=0). Then, it assigns a score according to the valid states of the world of web services.

Therefore, the matching between the request and a set of Web services can be measured quantitatively. The service has a high similarity score represents the service the most accurate for the request.

The following equation generalizes the comparison between the proposed request by the composition module and web services:

$$Sim(Req, SW) = Sim_{In/Out}(Req, SW) + Sim_{Pre/Effet}(SW) \quad (5)$$

Where

$$Sim_{In/Out}(Req, SW) = \begin{cases} 3 \ (Exact) & if \ InReq = InSW \\ & and \ OutReq = OutSW \\ 2 \ (plug\text{-}in) & if \ InReq \supset InSW \\ & and \ OutReq \subset OutSW \\ 1 \ (Subsumes) & if \ OutReq \supset OutSW \\ 0 \ (Fail) & if \ OutReq <> OutSW \end{cases}$$

(6)

$$Sim_{Pre/Effet}(SW) = \begin{cases} 2 & if \ Valid(Pre\text{-}Condition) \wedge Valid(Effet) \\ 1 & if \ Valid(Pre\text{-}Condition) \vee Valid(Effet) \\ 0 & if \ \neg Valid(Pre\text{-}Condition) \wedge \neg Valid(Effet) \end{cases}$$

(7)

With: InReq denotes the inputs of the request; OutReq denotes the outputs of the request; InSW denotes the inputs of the web service; OutSW denotes the outputs of the web service.

The architecture of this module is illustrated in the diagram below:

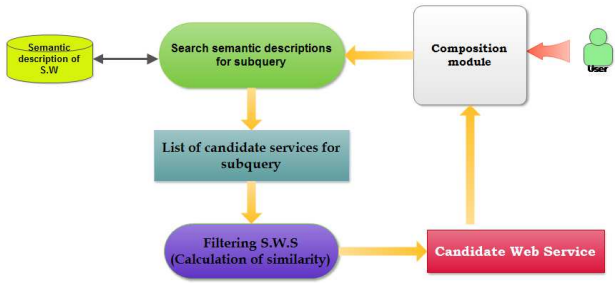

Fig. 3 Architecture of discovery module

**Discovery algorithms**

Let:
SW: a set of existing Web services in the directory.
SReq: the sub query.
SWF: selected Web service

**Algorithm 1: Module Discovery**
**Input: SW, SReq**
**Output: SWF**
**Taux_sim ← 0**
    For each S in SW do
        If Similarité (SReq,S) > Taux_**sim** Then
            Taux_sim ← Similarité (SReq,S)
**SWF ← S**
        End if
    End for
**Return (SWF)**
**End.**

**Algorithm 2: Semantic Similarity**
**Input : SW, SReq**
**Output :Taux_Sim**
**If** EntReq = EntSW **and** SortReq = SortSW **Then**
**Taux_Sim ← 3**
**Else if** EntReq ⊃ EntSW **and** SortReq ⊂ SortSW **Then**
    Taux_Sim ← 2
**Else if** SortReq ⊃ SortSW **Then**
**Taux_Sim ← 1**
**Else**
    Taux_Sim ← 0
**End if**
**If** Pre-ConditionSW = vrai **and** EffetSW = vrai **Then**
**Taux_Sim ← Taux_Sim+2**
**Else if** Pre-ConditionSW = vrai **or** EffetSW = vrai **Then**
    Taux_Sim ← Taux_Sim+ 1
**End if**
**Return (Taux_Sim)**
**End.**

## 4. Conclusion

In this paper, we proposed an approach for automated Web services composition based on directed graphs theory. In this approach, we proposed a formal method for describing Web services, then selecting and ordering the ones which satisfy the required inputs and outputs for the compound Web service.

In future work, we will be working on a module for the verification and validation of composition with respect to user needs. We will also be working on a model of semantic representation of web services in order to assess the degree of similarity or the possibility of interaction between Web services.